\documentclass[sigconf]{acmart}

\usepackage{rotating}
\usepackage{verbatim}
\usepackage{enumitem}
\usepackage{adjustbox}
\usepackage{multicol}
\usepackage{multirow}
\usepackage{fancyhdr}
\usepackage{fancyvrb}
\usepackage[linesnumbered,ruled,vlined]{algorithm2e}
\usepackage[normalem]{ulem}
\usepackage{algpseudocode}
\usepackage{wrapfig,lipsum,booktabs}
\usepackage{pifont}
\usepackage{tabularx}
\usepackage{mdframed}

\DontPrintSemicolon

\algrenewcommand{\algorithmiccomment}[1]{\hfill// #1}

\definecolor{comment}{rgb}{0,0.6,0}
\definecolor{dgray}{gray}{0.25}
\definecolor{strings}{RGB}{0,0,128}
\definecolor{backcolour}{rgb}{0.95,0.95,0.92}
\definecolor{keywords}{RGB}{127,0,85}
\definecolor{darkred}{RGB}{139,0,0}
\definecolor{darkyellow}{RGB}{204,204,0}
\definecolor{darkblue}{rgb}{0.0, 0.0, 0.55}
\definecolor{vividviolet}{rgb}{0.62, 0.0, 1.0}
\definecolor{fuchsia}{rgb}{1.0, 0.0, 1.0}
\definecolor{shockingpink}{rgb}{0.99, 0.06, 0.75}
\definecolor{darkBlue}{RGB}{0, 0, 205}

\newcommand{\cmark}{\text{\ding{51}}}
\newcommand{\xmark}{\text{\ding{55}}}

\newcommand{\josh}[1]{{\color{shockingpink}(Josh: #1)}}

\newcommand{\alfred}[1]{}

\newcommand{\rhl}[1]{}
\newcommand{\modify}[1]{#1}
\authorsaddresses{}
\mdfdefinestyle{AnswerRQ}{%
    linecolor=black,
    outerlinewidth=0.3pt,
    roundcorner=10pt,
    skipabove = 4pt,
    innertopmargin=5pt, 
    innerbottommargin=5pt, 
    innerrightmargin=5pt,
    innerleftmargin=5pt,
    backgroundcolor=gray!10!white
}

\setcopyright{none}
\renewcommand\footnotetextcopyrightpermission[1]{}
\begin{document}

\title{Towards Automated Driving Violation Cause Analysis in Scenario-Based Testing for Autonomous Driving Systems}

\author{{\rm Ziwen Wan}$^{}$\quad {\rm Yuqi Huai}$^{}$ \quad {\rm Yuntianyi Chen}$^{}$ \quad {\rm Joshua Garcia} $^{}$ \quad {\rm Qi Alfred Chen}\\
University of California, Irvine \ \ \ \{ziwenw8, yhuai, yuntianc, joshug4, alfchen\}@uci.edu}

\renewcommand{\shortauthors}{}
\newcommand{\outline}[1]{}

\newcommand{\todo}[1]{}

\newcommand{\ziwen}[1]{{\color{purple}(Ziwen: #1)}}

\newcommand{\cut}[1]{}

\settopmatter{printacmref=false, printfolios=false}
\setcopyright{None}


\begin{abstract}

The rapid advancement of Autonomous Vehicles (AVs), exemplified by companies like Waymo and Cruise offering 24/7 paid taxi services, highlights the paramount importance of ensuring AVs' compliance with various policies, such as safety regulations, traffic rules, and mission directives. Despite significant progress in the development of Autonomous Driving System (ADS) testing tools, there has been a notable absence of research on attributing the causes of driving violations.
Counterfactual causality analysis has emerged as a promising approach for identifying the root cause of program failures. While it has demonstrated effectiveness in pinpointing error-inducing inputs, its direct application to the AV context to determine which computation result, generated by which component, serves as the root cause poses a considerable challenge. A key obstacle lies in our inability to straightforwardly eliminate the influence of a specific internal message to establish the causal relationship between the output of each component and a system-level driving violation.

In this work, we propose a novel driving violation cause analysis (DVCA) tool. We design idealized component substitutes to enable counterfactual analysis of ADS components by leveraging the unique opportunity provided by the simulation. We evaluate our tool on a benchmark with real bugs and injected faults. The results show that our tool can achieve perfect component-level attribution accuracy (100\%) and almost (>98\%) perfect message-level accuracy. Our tool can reduce the debugging scope from hundreds of complicated interdependent messages to one single computation result generated by one component. 

\end{abstract}
\settopmatter{printfolios=true} 

\maketitle
\pagestyle{plain}

\section{Introduction}

The development of Autonomous Vehicles (AVs) has been remarkably swift in recent years. Notably, Waymo has received authorization to offer 24/7 paid taxi services in California~\cite{cali_taxi}. In this context, the significance of AV safety has grown exponentially, given that safety incidents can inflict significant damage to a company's reputation and finances, as illustrated by recent fatal accidents involving Uber and Tesla~\cite{uber_crash, tesla_crash_truck, tesla_crash_texas, tesla_mich}. Safety reports from leading companies, including Waymo, GM Cruise, Argo AI, and Pony.ai~\cite{waymo_safety_report, gmcruise_safety_report, argo_safety_report, ponyai_safety_report}, underscore the prioritization of safety as their foremost concern. Beyond safety, it's worth noting that AVs have also encountered unexpected immobilization issues in real-world scenarios~\cite{NRP_report}.

Detecting driving violations (e.g., safety violations, traffic rule violations, driving mission violations) before large-scale deployment is undeniably crucial, leading to significant advancements in Autonomous Driving System (ADS) testing research. A recent survey~\cite{testing_survey} reveals that over the past four years, more than 30 papers dedicated to ADS testing have been published in top-tier conferences and journals. Many ADS testing tools have demonstrated their ability to automatically discover driving violations, such as collisions, traffic rule violations, and failure to complete driving missions, using simulation. Surprisingly, despite the substantial volume of research in ADS testing, none have delved into the analysis of the root causes of driving violations in ADS.

Debugging within ADS presents substantial challenges. These systems consist of multiple intricate components, ranging from Deep Neural Network (DNN)-based object detection and prediction to programmed decision-making software. These components interact at a high frequency due to the real-time nature of the system. In a minute, thousands of messages are generated and transmitted in ADS, as observed in systems like Baidu Apollo and Autoware~\cite{apollo, autoware}.
Manual analysis of driving violations in such a complex environment is both expensive and non-scalable. Therefore, the need for an automated tool that can effectively and efficiently analyze the root causes of driving violations is paramount, given the criticality of safety in ADS. Imagine having a tool that can promptly identify the message (output of an ADS component) responsible for a violation immediately after a bug is detected by a testing tool. This would enable rapid assignment of the bug to the respective developer, freeing them from the burden of comprehending intricate component interactions to determine the cause of driving violations. Moreover, armed with the knowledge of which message is triggering the violation, understanding and rectifying the bug would be significantly less time-consuming.

Therefore, the focus of this paper is to address the following question: How can we develop an effective and efficient automated tool for attributing the causes of driving violations? As outlined earlier, this tool should accomplish two primary attribution goals: (1) identifying the responsible ADS component for the violation, and (2) pinpointing the specific message that induces the violation.

However, the design of such a tool presents inherent challenges. A notable hurdle is that driving violations, such as collisions, are defined at the system level, which represents the cumulative outcomes of component interactions. Establishing a direct link between a system-level driving violation and the output of an individual component is not straightforward. Recent research~\cite{jung2022swarmflawfinder, jung2021swarmbug, von2021icra} on the analysis of Cyber-Physical Systems (CPS) software has suggested that counterfactual causality could offer a promising approach. The fundamental concept here is that by selectively removing certain circumstances (e.g., obstacles in the surrounding environment) and re-running tests, we can ascertain their influence. Applied to our context, if the driving violation no longer occurs after the removal of specific components or messages, it strongly indicates that the removed elements are the root cause of the violation.

While the concept of removing components or messages and re-running tests is promising, it is not a straightforward process when applied in an ADS context to attribute the root cause of driving violations. 
Eliminating the impact of an output message and conducting re-testing is not as straightforward as removing a physical obstacle. The primary challenge arises because this output message is promptly utilized as input for another component. Merely removing the message will inevitably disrupt the proper functioning of the other component.

To address the challenges outlined earlier, we introduce the concept of idealized component substitutes for re-testing within the ADS context. Leveraging the advantages offered by simulation, we generate expected outputs for each component. Consequently, we can selectively remove specific components or messages to discern the counterfactual relationship between the system-level testing oracle (driving violation) and component outputs. We present a novel framework that capitalizes on the simulator's unique capabilities to facilitate causality analysis in ADS scenario-based testing.


To further evaluate the effectiveness of our tool, we created a benchmark with real bugs. Our tool can achieve perfect accuracy (100\%) on both component-level and message-level cause attribution. We further evaluated our tool with injected faults to ensure the comprehensiveness of the evaluation. Overall, our tool can reduce the debugging scope to one output message correctly on almost (>98\%) all the cases within 20 minutes.

In summary, our work makes the following major contributions:

\begin{itemize}[leftmargin=*,nosep]

   \item We are the first to study automated driving violation cause attribution within the domain of ADS. This endeavor is deemed essential and significant, considering the paramount importance of safety in ADS. We have framed the problem and recognized the distinct challenges specific to this context. 
    \item We design and implement an automated framework for ADS driving violation cause attribution. We utilize the unique opportunity in the simulation environment to enable counterfactual causality analysis in the ADS context.
    \item \modify{We evaluate our tool on a benchmark with real bugs from popular open-source industry-grade ADS software. The results indicate that our tool can achieve perfect accuracy on both component level and message level and at least 98.5\% reduction rate. To ensure the evaluation comprehensiveness, we also evaluate it on injected faults. The results show that our tool can effectively identify the violation-inducing message from diverse components.}
\end{itemize}

\section{Background and Motivation}
\begin{figure*}[t]
    \centering
    \includegraphics[width=\linewidth]{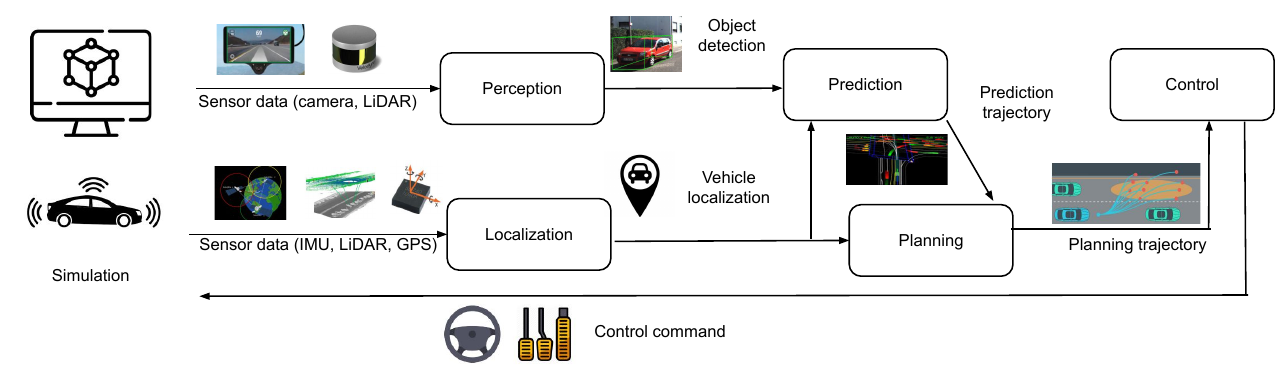}
    \caption{\modify{Overview of representative ADS systems and the interactions between its internal components.} \alfred{the text size in the figure is best to be similar and a little bit smaller to the main paper text size}}
    \label{fig:ad_framework}
\end{figure*}

\subsection{ADS and Typical ADS Components}

Autonomous Vehicles (AVs), or self-driving cars, are already a reality in our daily life --- over 60 companies are testing such systems on public roads today~\cite{dmv-permit-autox}, and quite a few already in the commercial stage, with notable examples of robo-taxis~\cite{waymo-one, baidu_driverless_robotaxi}, buses~\cite{baidu_bus_service, uk_bus}, trucks~\cite{tusimple-truck, aurora_truck}, and delivery vehicles~\cite{nuro_charge_delivery_service}, not to mention the millions of Autopilot-equipped Tesla cars~\cite{tesla_sold_2_million_cars}.

To achieve safe driving autonomy in complex and dynamic driving environments, ADS in the AV industry today typically adopts a \textit{multi-component architecture}~\cite{waymo_safety_report, aurora_safety_report, gmcruise_safety_report, apollo, autoware}, where each component is in charge of a relatively self-contained AI functionality and frequently interacts with each other to realize an end-to-end ADS mission. The input and output of each component are transmitted via \textit{messages} in a robotic middleware such as ROS~\cite{quigley2009ros}.     

The components of ADS can be classified into three stages: \textbf{1. Sensing} components refer to the components that collect information and extract knowledge from the environment. Perception detects the position, size, and velocity of surrounding traffic objects from the sensor data, the prediction component generates predicted trajectories of surrounding traffic objects, and the localization component estimates the current position of the ego vehicle from various sensor data.
\textbf{2. Planning} refers to making a high-level driving decision based on the informatio gained from the sensing stage. \textbf{3. Control} refers to executing the high-level decision from the planning step with low-level vehicle control commands. \alfred{should still try to remove the orphans/widows like this}

\modify{During the execution, each component will store the subscribed messages, and the main logic of each component will be executed driven by the clock or a certain type of message. Eventually, the output of each execution will be sent to other components as an output message via the APIs provided by the middleware. We can use the following triad to define an execution of a certain component:

\begin{equation}
    e \;:= \;<\, m_{input}, \, m_{output}>
\end{equation}
where $m_{input}$ is the message containing all inputs of the component, $m_{output}$ is the output message of the program execution.

During the execution of a driving scenario, each component will continuously collect and transmit messages such that the whole AD vehicle can understand and react to the complicated environment in real-time. Thus, the executions of each component will naturally form a vector where the index reflects the execution order. We define the execution vectors of each component as $E^{perc}$, $E^{pred}, E^{plan}, E^{cont}, E^{loca}$ for perception, prediction, planning, control, and localization component respectively. Furthermore, the whole execution of the ADS system could be represented as a list of vectors:
\begin{equation}
\mathbb{E} = \left\{
\begin{aligned}
< & e^{perc}_1 & e^{perc}_2 & e^{perc}_3 & ... & e^{perc}_{|E_{prec}|}> \\
< & e^{pred}_1 & e^{pred}_2 & e^{pred}_3 & ... & e^{pred}_{|E_{pred}|}> \\
< & e^{plan}_1 & e^{plan}_2 & e^{plan}_3 & ... & e^{plan}_{|E_{plan}|}> \\
< & e^{cont}_1 & e^{cont}_2 & e^{cont}_3 & ... & e^{cont}_{|E_{cont}|}> \\
< & e^{loca}_1 & e^{loca}_2 & e^{loca}_3 & ... & e^{loca}_{|E_{loca}|}>\\
\end{aligned}
\right\}
\end{equation} 
where $e_i^X$ means the execution from component $X$ with $i-th$ smallest timestamp. Each row is the execution of the same component sorted with the timestamp. Each component may have a different number of executions even during the same period, which means the number of elements on each row may vary. 
}

\subsection{Scenario-based Testing for ADS}

Considering that cars are heavy, fast-moving, and operating in public spaces, safety has always been the most important concern for AV developers and policy-makers. To address this, end-to-end system-level safety testing has been one of the most critical steps in real-world ADS development and deployment. Road testing is the most natural approach for such safety testing, but due to the large variety of corner-case scenarios and their rarity in naturalist driving, such testing has been widely recognized to be extremely inefficient. For example, it is estimated that an ADS must drive hundreds of millions of miles to validate the safety at the level of human-driven vehicles~\cite{kalra2016driving}, which makes it intractable in practice.

To address this, \textit{scenario-based testing} has been widely adopted in the AV industry, where testers can simulate rare corner-case driving scenarios, typically in a high-fidelity simulator, to test a given ADS with flexibility and scalability. For example, Waymo has reported more than 15 billion miles of such virtual testing of their ADS in their proprietary high-fidelity simulator~\cite{waymo_safety_report}; Aurora claimed that such a method enables them to perform 2.27 million unprotected left turns in simulation before even attempting one in the real world.~\cite{aurora_scale_simulation}. Given such a large volume of testing instances, manual analysis of the root cause is not scalable anymore and it is necessary to have an automated root cause analysis approach. However, how to automatically and systematically investigate the root cause of failed tests is not discussed in existing work.

\cut{
\subsection{Testing AD Systems}

Based on the prior survey~\cite{lou2021testing}, there are three commonly used testing methods for AD systems: unit testing, field operational testing, and simulation testing. The developers can manually label and clip driving records from simulation or field operational tests to construct unit tests for control logic and DNN models~\cite{lou2021testing}. However, to verify the safety property of an entire system, we still need to rely on field operational testing or simulation testing. Simulation testing has been widely deployed by AD companies. Even though the environment and vehicle dynamics may be different in the physical world, it is an irreplaceable approach due to its low cost, high efficiency, and the capability to turn one single physical-world case into thousands of opportunities to practice and improve AD software~\cite{waymo_safety_report}. More background information in this line of research is included in a recent survey~\cite{testing_survey}. Among them, a few recent works~\cite{autofuzz, li2020av} have built heuristic-based algorithms to boost the bug discovery process, and they can find safety-violation scenarios for popular open-source multi-component AD frameworks (e.g., Apollo, Autoware~\cite{apollo, autoware}). 

We also want to highlight a few key differences that make AD simulation testing particularly different from traditional software testing: (1) The output of AD simulation testing is a scenario configuration, which is the input to the simulator instead of inputs to the program under test. (2) The oracle is defined at the system level (e.g., collision), which is a joint effect caused by all components of AD software and simulator. \josh{Please describe why you are highlighting these differences.}

}

\subsection{Motivation}
\begin{figure*}[t]
      \centering
          \includegraphics[width=1.\linewidth]{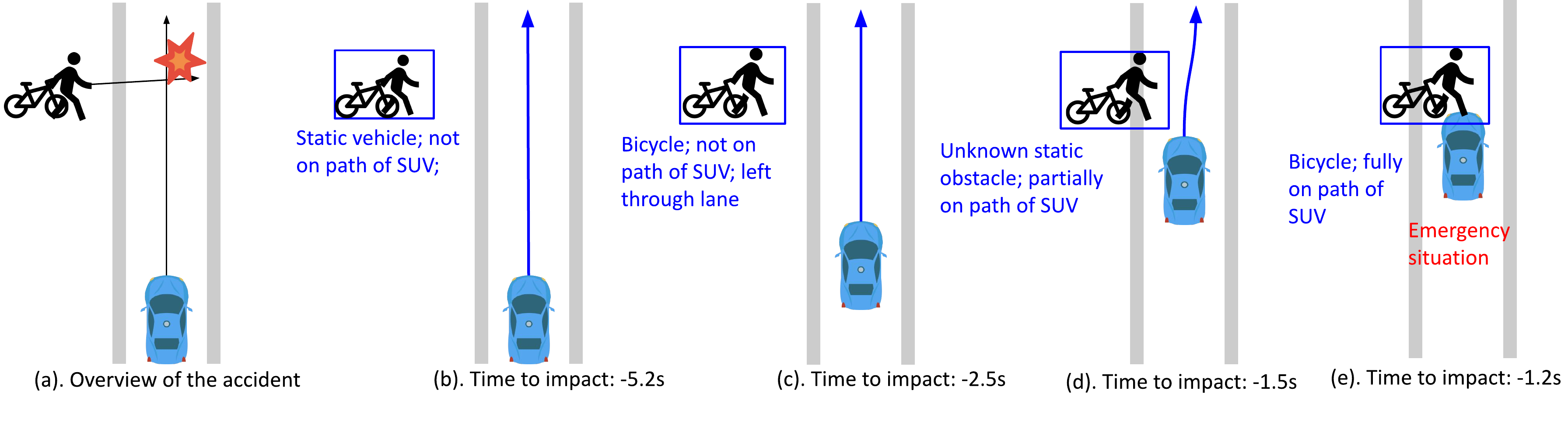}  
	\caption{\modify{Illustration of collision between vehicle controlled by the AD software and the pedestrian. (a). shows the moving trajectories of both the vehicle and the pedestrian in the accident. (b)-(e) shows the autonomous vehicle's classification and prediction results of the pedestrian and its corresponding decision. Figures are generated based on the description in NTSB's investigation report~\cite{ntsb_report}. \alfred{text size in the figure too small; see figure 1 comments for best practice suggestion}} }
\label{fig:moti_example}
\end{figure*}

We introduce a motivating example to demonstrate how isolating failure-inducing input is different in the AD context compared with general software debugging. This example is a real fatal collision between a vehicle operated by AD software and a pedestrian pushing a bicycle by her side. Being the first pedestrian fatality involving an AD vehicle, this is one of the most high-profile AD collisions in the world. The report of the National Transportation Safety Board (NTSB)~\cite{ntsb_report} explains how it happened. As shown in Fig.~\ref{fig:moti_example} (a), the vehicle had a collision with a pedestrian when she was crossing the road. 5.2 seconds before the collision (Fig.~\ref{fig:moti_example} (b)), the AD vehicle detected the pedestrian for the first time and mistakenly recognized the pedestrian as a static vehicle. When it was 1.5 seconds before the collision, the pedestrian had 
entered AD vehicle's lane, and she was again mistakenly classified as some unknown static obstacle.
The AD vehicle generates a planning trajectory to avoid the collision with the static obstacle. However, since it mistakenly thought the object was static, that collision-avoidance planning trajectory could not really avoid the collision once the pedestrian moved closer to the AD vehicle. Finally, when it was 1.2 seconds before the collision, the AD vehicle realized that the collision might happen and claimed an emergency situation. 
The report claims that the alternating classification cleared the tracking history of the pedestrian, which led to an incorrect prediction of the pedestrian's trajectory, and ultimately caused the fatal collision.


Understanding the root cause of this accident is far from trivial. Just by looking at the accident description (Figure ~\ref{fig:moti_example}.a), many possibilities exist. The pedestrian was not detected, the pedestrian's tendency towards the AD vehicle was mistakenly ignored, the decision was too aggressive to avoid the collision or even the control command may have contained an error such that the AD vehicle was out of the control of AD software. If our step can go one step further and we can see how AD vehicle's internal decision information (Figure ~\ref{fig:moti_example}.b-e), we can exclude some of the assumptions. For example, we can confirm that the pedestrian is detected and the root cause should not be the control command error since the vehicle correctly follows the planning trajectories. However, we still can not confidently conclude the root cause. Even though we know the prediction trajectory is wrong, we can not tell whether this error is an inherent error inside the AD's prediction component or caused by the misclassification results from the perception component. Furthermore, there is also the possibility that the planning component was not ready to handle the pedestrians who 
are jaywalking.
A natural way to further understand the cause of this accident is to raise some what-if questions: "What if the perception component does not misclassify the pedestrian?" or "What if the prediction component always generates the correct prediction trajectory?". In this paper, we will demonstrate how we understand the root cause of AV accidents by asking and answering the what-if questions. 

Another motivating factor for our work is the automation of the entire process. The components of ADS are running at a high frequency (in Apollo~\cite{apollo}, each component is usually operated with at least 10Hz and some components are even operated with 100Hz) due to the requirement of quick response in such safety-critical systems. More than 100 messages could have been generated within one second. Let only each message also contain high-dimension data. For example, one single planning message could contain more than one thousand numerical values, since it has to include all detailed information for the planning trajectory. Given the large volume of data contained in a driving trace, the need for iterative reduction of the scope of debugging through the continuous exploration of "what-if" questions becomes essential. This underscores the necessity of an automated approach. 

\section{Simulation-based Testing Specification}

To clearly describe the problem and our solution, we present a formal specification of the simulation-based testing. We follow common specifications in recent works on AD software testing~\cite{huai2023doppelganger}. Our specification is reduced from the original version~\cite{huai2023doppelganger} since only one vehicle is controlled by the AD software in our context.

\textbf{Definition 1 (testing scenario).} \textit{A scenario $S = <A_{init}, A_{dest}, \mathbb{O},\\\mathbb{T}, t^S>$ is a tuple where: $A_{init}$ and $A_{dest}$ is the AD vehicle's initial location and destination location which will be represented under a certain coordinate (e.g., UTM coordinate); $\mathbb{O} = \{O_0, O_1, ..., O_{|\mathbb{O}|}\}$ is a finite set of traffic objects where $|\mathbb{O}|\geq 0$; $\mathbb{T}$ is the traffic signal configuration that includes color of traffic signals at each moment during the driving; $t^S$ is the maximum allowed duration for the testing scenario. }

\textbf{Definition 2 (traffic object).} \textit{A traffic object $O = <\mathbb{W}^O, type^O, \\size^O>$} is a tuple where: $\mathbb{W}^O = <w_0^O, w_1^O, ..., w_{|\mathbb{W}^O|}^O>$ is a sequence of waypoints (Definition 3) that the traffic object will follow at each time; $type^O$ and $size^O$ are the type and size of the traffic object. The traffic object could be a walking/standing pedestrian, a static/moving vehicle, or even a static object such as a traffic cone. 

\textbf{Definition 3 (waypoint).}
\textit{A waypoint $w$ is represented by a tuple $<p, v, a, t>$}. $p$, $v$, and $a$ are all vectors representing the AD vehicle or the position, velocity, and acceleration of any traffic object at each time $t$. It is noteworthy that even if a traffic object remains stationary, it will still have a sequence of waypoints to represent its potential movement trajectories. In such cases, the object's position will remain unchanged and both its velocity and acceleration vectors will be consistently zero throughout the duration.

\modify{\textbf{Definition 4 (\textit{rtest}).}
\textit{The function $rtest$ determines for a program run given whether some specific violation is detected.} It can be represented as the following equation:
\begin{equation}
    rtest(S, \; AD\_Software, \; V) = \cmark / \xmark, \; \mathbb{A}, \; \mathbb{E}
\end{equation}
where $S$ is a testing scenario (Definition 1), $AD\_Software$ is a certain version of the AD software, $V$ is a set of traffic violations we want to detect, $\cmark / \xmark $ indicating whether any violation is detected ($\xmark$) or not ($\cmark$), $\mathbb{A} = <A_0, A_1, A_2, ... A_{|\mathbb{A}|}>$ is a set of waypoints representing the position, velocity, and acceleration of the ego vehicle at each moment, and $\mathbb{E}$ (detailed in Definition 5) is the collection of component executions. For the remaining part of the paper, we will use a simplified version of the function when the AD software and the types of violations are fixed: $rtest$ : 
\begin{equation}
    rtest(S) = \cmark / \xmark
\end{equation}
In remaining of the paper, the software and the violations used are fixed, and such a simplified expression will not lead to confusion.}

\textbf{Definition 5 (message).} 
\textit{We denote $\mathbb{M}$ as the collection of all messages generated during the simulation test. For ease of representation, we denote $\mathbb{M}[module][i]$ as the $i$th message generated from the given $module \in \{ \rm Perception,\rm Prediction, Planning, Control, Localization\}$.} We also denote $\mathbb{M}[module][:]$ as all messages generated from a certain module and $\mathbb{M}[module][i:-1]$ as all the messages generated from $module$ and with index $\geq i$.

We further use temporal logic to represent the violations we want to detect during the simulation-based testing. The definition can be customized based on the different operation domains and testing standards of AD vehicles. We only introduce the most basic violations in our design to demonstrate the generality of our tool.

\textbf{Definition 6 (\modify{safe distance violation}).} 
\modify{The paramount rule for an Autonomous Driving (AD) vehicle is maintaining a safe distance from other vehicles during normal driving conditions. This rule stands as the utmost priority for AD vehicles since any violation of the safety distance could potentially lead to a collision that causes substantial harm.}
\textit{The specification must always be adhered to, and any violation of rule will be regarded as either a collision or hazardous driving behavior, depending on whether an actual collision occurs. }
\begin{equation}
    \forall t \;\forall O,\quad dist(bbox(A, t), bbox(O, t)) \geq c
\end{equation}

\textit{where $dist(\cdot)$ function is defined as the minimal distance between two bounding boxes, $bbox$ function is a function generated by the corresponding box for each traffic object or the ego vehicle at time $t$, and $c$ is a constant value representing the minimal safe distance between ego vehicle and other traffic object. Any violation of the above-mentioned rule is a safe distance violation.}

\textbf{Definition 7 (driving mission violation)}.  \textit{We also want to check whether the AD vehicle can finish the driving task (i.e., reaching the given destination). The specification is described as follows and any violation of such specification is a driving task violation:}
\begin{equation}
    samePosition(A_{|\mathbb{A}|}.p, A_{dest}) = True
\end{equation}
\textit{where function $samePosition$ means whether the two points are in the same position (a domain-specific function is needed since the numerical values cannot be precisely the same). We compare the final location of the ego vehicle $A_{|\mathbb{A}|}.p$ with the given destination $A_{dest}$. }

\textbf{Definition 8 \modify{(speeding limit violation)}}. \textit{ Another safety-critical violation is the speed limit violation. The specification is described as follows and any violation of such specification is a speeding violation:}
\begin{equation}
    \forall t, \quad speedEgo(A, t)\leq speedLimit 
\end{equation}
\textit{where $speedEgo$ is a function to query the speed of ego vehicle at time $t$, and $speedLimit$ is the speed limit on the current lane.}
\section{Problem Formulation and Challenges}
\label{sec:formulation}

\subsection{Problem Formulation: Driving Violation Cause Analysis (DVCA)}
\label{formulation:formulation}


As the first work towards automating the driving violation cause analysis problem in scenario-based testing, or the DVCA problem for short, we first systematically formulate its design requirements and goal as follows:

\textbf{Scope of violations.} Our tool's design has considered three types of driving violations: safety violations, traffic rule violations, and driving mission violations. 

The most crucial category of driving violation pertains to safety violations, which undoubtedly have the potential to cause substantial harm within ADS. The collision involving Uber's autonomous vehicle~\cite{ntsb_report} resulted in the loss of one life and inflicted substantial financial damage on Uber. In our characterization, safety violations typically encompass collisions or hazardous closeness between vehicles, pedestrians, or any other objects within the driving scenario. Additionally, we consider traffic rule violations, which can also have significant safety implications. Traffic rule violations encompass behaviors that blatantly contravene traffic regulations, such as exceeding speed limits or deviating from designated lanes.

While safety violations and traffic rule violations cannot be tolerated, it's important to highlight that our tool also facilitates the understanding of the root causes behind driving mission violations. These instances involve autonomous vehicles unexpectedly failing to carry out their designated driving missions, a matter that has gained considerable attention in both the research community and within AD vehicle companies recently~\cite{planfuzz, NRP_report}. For example, consider the scenario in which protestors placed traffic cones in front of a self-driving car, resulting in the vehicle's immobilization. Driving mission violations encompass situations such as the failure to proceed through an intersection without safety concerns or the inability to reach the intended destination. component

\modify{\textbf{Gap between scenario-based testing oracles and execution-level testing oracles.} After finding certain driving violations, a natural step is to localize the fault inside the AD software such that the developers can quickly improve the robustness and safety of the critical system. However, on one side, the testing oracle (driving violation) is defined in a driving context. Once given an execution matrix, we could only assign a boolean value to this whole-system execution trace. The mapping between the executions and the pass/fail results is a many-to-one mapping: $\mathbb{E} \rightarrow \{\cmark, \xmark\}$. On the other side, fault localization techniques require more fine-tuned pass/fail labeling on one execution: $e \rightarrow \{\cmark, \xmark\}$, which is a one-to-one mapping from one execution to one pass/fail label. 

Many advanced fault localization techniques~\cite{wong2016survey} (e.g., spectrum-based techniques, data mining-based techniques, statistics-based techniques) require the existence of testing oracles, which means given one single execution instance, we can precisely tell whether this execution is a pass or failed test case. 
But in a driving trace including a driving violation, the violation might be caused by a very limited number of faulty executions and the remaining ones should be marked as pass. If we directly use the scenario-based testing oracles, it will certainly include misleading information due to the one-to-one pass/fail labeling. To overcome this gap, we propose to use cause analysis to identify faulty executions inside a driving trace. Thus, our cause analysis can be considered as a process to extend the many-to-one scenario-based testing oracle to more fine-grained many-to-many testing oracles. Our process is expected to provide such mapping information:

\begin{align*}
    CA\left(\left\{\begin{aligned}
e^{perc}_1 & e^{perc}_2 & e^{perc}_3 & ... & e^{perc}_{|E_{prec}|} \\
e^{pred}_1 & e^{pred}_2 & e^{pred}_3 & ... & e^{pred}_{|E_{pred}|} \\
e^{plan}_1 & e^{plan}_2 & e^{plan}_3 & ... & e^{plan}_{|E_{plan}|} \\
e^{cont}_1 & e^{cont}_2 & e^{cont}_3 & ... & e^{cont}_{|E_{cont}|} \\
e^{loca}_1 & e^{loca}_2 & e^{loca}_3 & ... & e^{loca}_{|E_{loca}|} \\
\end{aligned}\right\} \right) = \\
\left\{
\begin{aligned}
?/\xmark/\cmark \quad &  ?/\xmark/\cmark & ?/\xmark/\cmark & ... & ?/\xmark/\cmark \\
?/\xmark/\cmark \quad &  ?/\xmark/\cmark & ?/\xmark/\cmark & ... & ?/\xmark/\cmark \\
?/\xmark/\cmark \quad &  ?/\xmark/\cmark & ?/\xmark/\cmark & ... & ?/\xmark/\cmark \\
?/\xmark/\cmark \quad &  ?/\xmark/\cmark & ?/\xmark/\cmark & ... & ?/\xmark/\cmark \\
?/\xmark/\cmark \quad &  ?/\xmark/\cmark & ?/\xmark/\cmark & ... & ?/\xmark/\cmark \\
\end{aligned}
\right\}
\end{align*}

where $CA$ is the cause analysis function. $?$, $\xmark$, $\cmark$ represents that this execution is unresolvable, and is or is not the root cause of the driving violation on the scenario level. With such many-to-many mapping, we can naturally break it down to get the one-to-one mapping for each single execution instance.
}

\textbf{Design goal.} Specific to the context of ADS and scenario-based testing, we identify two concrete DVCA design goals that can most generally benefit ADS developers in practice:


\begin{itemize}[leftmargin=*,nosep]
   \item \textbf{Component-level cause attribution.} Given an ADS system-level driving violation (e.g., crashing into road objects such as cars and pedestrians), to facilitate debugging, it is highly desired to quickly know which ADS component is most likely responsible for such an end-to-end system-level consequence. This can dramatically reduce the debugging scope from simultaneously tracking multiple components and their behavioral inter-dependencies to tracking only one single component. Moreover, since in industry ADS development teams, the components of different functionality types (e.g., perception, prediction, planning) are typically developed by different sub-teams, such component-level cause attribution can also best facilitate the debugging task allocations in such common system development models.
    
    \item \textbf{Message-level cause attribution.} Once the developer narrows down the specific ADS component responsible for a violation, a challenging debugging task remains. As ADS operates in real-time, each ADS component receives input messages at a frequency of at least 10Hz. Given that each driving simulation test case lasts 1-2 minutes~\cite{autofuzz, li2020av}, even at the individual ADS component level, there are between 600-1200 component executions to diagnose in order to understand which one(s) led to the system-level driving violation. Consequently, after component-level attribution, it is ideal to automatically identify which input/output message(s) from the violation-inducing component is the root cause of the violation. It's important to note that in the context of ADS, message-level cause attribution holds particular utility for developers, especially when dealing with data-driven AI models, as identifying violation-inducing message(s) can be instrumental in model correction. For instance, this can involve retraining with the violation-inducing model input and correctly labeled output data. Our objective is to pinpoint a output message that developers should prioritize, thus streamlining debugging efforts.

\end{itemize}

\subsection{Design Challenges}
\label{formulation:challenge}

Although there are various prior designs on automated cause analysis for different software failure symptoms in software systems~\cite{liu2021microhecl, malik2013automatic, murali2021scalable, johnson2020causal}, none of them can be directly applied to our problem settings due to two main problem-specific design challenges:

\textbf{C1: Lack of easily-inferrable causal relationship between component-level errors and system-level consequences in ADS.} To achieve both component- and message-level cause attribution, we need to causally associate a given ADS system-level driving violation (e.g., vehicle collision) with a certain component-level error (e.g., an object misdetection in a particular frame in Perception). However, such an association is far from trivial in both directions. First, given the system-level violation, we cannot easily infer which component is the cause, since in the ADS context it is very common that many components in the driving decision-making pipeline can all generate errors that can cause exactly the same system-level violation symptom. For example, the same system-level violation symptom of a front-car collision can be caused by a Perception-only error that does not detect the front car; a Prediction-only error that does not correctly predict a sharp-braking behavior of the front car; a Planning-only error that does not brake accordingly when the front car has sharp braking; or a Control-only error that does not apply brake timely.

Second, from a component-level error (e.g., object misdetection in Perception), it has been widely recognized in the CPS area that such errors do not necessarily lead to system-level effects such as collisions~\cite{dreossi2019compositional, seshia2020semantic}. For ADS, this is especially true due to the high end-to-end system-level complexity and closed-loop control dynamics, which can explicitly or implicitly create fault-tolerant effects for the ADS component-level errors. Various such examples have already been discovered, e.g., when the object detection model error is at a far distance for automatic emergency braking systems~\cite{dreossi2019compositional, seshia2020semantic}, or such errors can be effectively tolerated by downstream ADS components such as object tracking~\cite{jia2020fooling}. Thus, although some prior works have designed component-level fault detection for robotic systems, e.g., using model-based and data-driven methods~\cite{khalastchi2018fault, yamaguchi2020specification}, they are not directly usable to solve the DVCA problem since even if they are able to detect faults, they cannot establish the causal relationship with the given system-level driving violation, not to mention the need to define effective component-level oracles in the first place, which is still a largely non-trivial research question.

\textbf{C2: Absence of novel framework designs for automated debugging in scenario-based testing.} A common assumption in software fault localization is that "the execution of a specific program is determined by a number of circumstances"~\cite{zeller2002simplifying}. Those circumstances could be the program code, code from storage, program inputs, and so on. In our cause attribution context, we regard each output message from each component as a circumstance and aim to ascertain which "circumstance" leads to the violations during the whole test. While this concept may seem akin to traditional fault localization, a key challenge in our problem lies in making each "factor" adjustable. It's relatively straightforward to nullify the effect of a factor if it constitutes part of the program inputs, simply by removing it. However, when we encounter a scenario where a factor is the output of a specific component, subsequently employed as inputs to other components to control the vehicle, the question arises: How can we manipulate or alter this particular factor effectively?

\cut{
\subsection{Application Scenario} 
\label{formulation:application_scenario}
The main scope of our cause localization is to understand the cause behind the collision incidents discovered in simulation testing. Additionally, we argue that the proposed techniques can also be applied for cause localization in field operational testing or even post-accident analysis via reconstructing the testing scenario in simulation. Waymo claimed that they have reviewed data from crash databases and naturalistic driving studies to
identify other possible collision scenarios and develop tests in simulation~\cite{waymo_safety_report}. Moreover, simulation has been used to reconstruct and analyze physical-world collision incidents~\cite{ubercrashreport}. Prior interviews with AD developers also justify this, as they mention that they can replay sensor data and control commands in simulation for regression testing~\cite{lou2021testing}. Thus, we believe system failures happening in the real world can also be analyzed in simulation environments. 
We focus on collision incidents since they are the most critical and severe system failures of AD systems. These types of incidents endanger human safety and are critical risks after the deployment of AD systems~\cite{uber_acquire, ubercrashreport}.

The types of causes this work focuses on include program implementation bugs or imperfections of the machine learning models used for object detection, prediction, and other tasks in AD software. Prior work has demonstrated that the implementation bugs may happen in different components in AD systems~\cite{garcia2020comprehensive}. Also, imperfect detection results are natural for machine learning models and a few real-world crash cases have demonstrated that such imperfect results can lead to severe collisions~\cite{ubercrashreport, tesla_crash_texas, tesla_crash_truck}.

Also, in this work, we focus on the localization of bugs that cause incorrectness of component outputs. Even though bugs that cause hanging, software crashes, or even delays can also lead to driving violations in such real-time systems, those bugs can be easily detected by monitoring the program's behavior and execution frequency. 

\subsection{Cause Localization: Needs and Design Requirements}
\label{formulation:needs}

Even though existing scenario-based testing tools have shown great promise in discovering driving-violation scenarios, to the best of our knowledge, there are no existing works that can analyze the cause of driving violations in multi-component AD systems automatically. 

\textbf{C1. Heterogeneous software implementation across components.} As prior work~\cite{lou2021testing} has discovered via interviews with AD developers, AD systems often use DNNs in tandem with explicitly programmed components, which poses unique challenges for software testing. Such heterogeneous software implementations, which we refer to as \textit{hybrid DNN+programmed (HDNNP)} software, pose new challenges in analyzing the cause of driving violations and potentially fixing bugs. There already exists a long line of research~\cite{wong2016survey} aiming to localize faults in traditional software. In newer domains, researchers have made significant progress in localizing and repairing DNN models \josh{missing citation}. However, we are not aware of any work that can combine existing techniques to support fault localization in an HDNNP software system.

\textbf{C2. Lack of component-level failed test cases due to usage of system-level oracles.} The current scenario-based testing is designed to test the safety of system behavior during the whole scenario and the oracle is defined at the system level (e.g., collision, lane violation,  which depends on the outputs of all components in AD software and simulator). However, each component will be executed multiple times during the simulation process. For example, the Perception and Prediction pipeline will be executed once a new sensor frame is published (the frequency is usually 10Hz in Apollo). Discovery of a driving violation does not necessarily mean all the executions of the program under test are wrong and due to system-level oracles, we can not identify the failed test cases of each specific component. However, the software fault localization techniques require failed test cases. As a result, such a system-level oracle further prevents us from applying existing software fault localization techniques to analyze the root cause of system-level driving violations. 
\textbf{}

Thus, we aim to design an automated tool that can provide more fine-grained information on the cause of driving violations. We envision this tool as a bridge to mitigate the gap between the safety-violation scenario discovered by the scenario-based testing tools and the requirements of existing software fault localization tools. Our scope of cause localization is focused on two dimensions: component-level localization and vehicle dynamic state-level localization. To address \textbf{C1}, our tool should be able to tell which component is faulty and the cause of a driving violation. Thus, the developers can further apply traditional software fault localization tools or DNN fault localization for logic-based components and DNN components accordingly. To address \textbf{C2}, our tool should be able to identify the ego vehicle dynamic state (e.g., position, velocity, acceleration) when the error happens. We choose to use vehicle dynamic state as the fine-grained information since such information is consistent across different executions, which would be demonstrated in \S (\todo{Add cross ref to design section}). With such cause localization information, our tool can serve as a bridge between the results of scenario-based testing tools and the requirements of existing software fault localization tools and it's the very first but necessary step towards analyzing and repairing the correctness bugs in AD software.

\todo{Consider expressing the problem with a more formal expression}
}
\section{Design}
\label{sec:design}


\begin{figure*}[t]
    \centering
    \includegraphics[width=\linewidth]{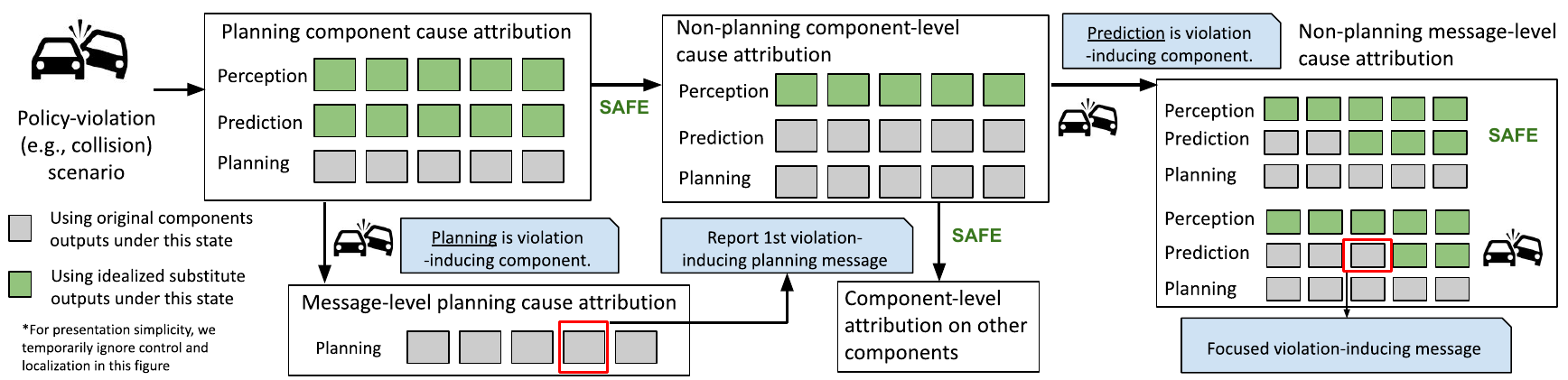}
    \caption{Illustration of working pipeline of our DVCA tool when analyzing a prediction-induced fault under a collision scenario. }
    \label{fig:PVCA_overview}
\end{figure*}

\subsection{Design Overview}
\label{design:overview}

In this section, we will present our high-level solution idea and an overview of our concrete automated DVCA tool design.

\textbf{Key solution principle: Counterfactual causality.} To address \textbf{C1}, our key idea is built upon the principle of \textit{counterfactual causality}~\cite{lewis2013counterfactuals} to establish a provable causal relationship between ADS system-level driving violation and component-level errors. In particular, it infers the causality via f counterfactual conditionals: ``If X had not occurred, Y would not have occurred''. Specifically, in our context of cause attribution in ADS, if a certain set of messages is not executed in the simulation testing, and the collision does not happen anymore, then the collision is causally dependent on the fault in that specific component.

\textbf{Definition 9 (delta test).} \textit{Following our counterfactual causality design, we introduce a new function $dtest$ (delta test), which aims to exclude the potential violation-inducing effects in a subset of output messages. More intuitively, we raise a question to the simulation testing tool, "What if those messages all perform correctly, what will happen, and whether the violation still occur?" The $dtest$ function is defined as $\mathcal{P} (\mathbb{M})\rightarrow \{\cmark, \xmark \}$, where $\mathcal{P} (\mathbb{M})$ is the power set of $\mathbb{M}$. Given each subset of $\; \mathbb{M}$, $dtest$ function will tell whether the violation still happens or not under a given testing scenario.}

\textbf{System inputs and outputs.} Following the idea of counterfactual causality, we start to formulate our problem inputs and outputs.  The \textbf{inputs} to our system is a driving violation scenario $S_{viol}$ where \begin{equation}
    rtest(S_{viol}) = \xmark \quad , which\; also\; means \quad dtest(\emptyset) = \xmark
\end{equation}

\textbf{Definition 10 (violation-inducing component)}. \textit{We define the violation-inducing component as the root cause of the violation on the component level. This is one of the outputs of our tool. One necessary condition for a violation-inducing component $component_{VI}$ is that $dtest(\mathbb{M}[component_{VI}][:]) = \cmark$, meaning that at least after removing the potential faulty effects from component $component_{VI}$, the violation no longer occurs.}

Another output from our tool is a specific violation-inducing message that we target. We treat planning components and non-planning components separately due to system design constraints, which will be elaborated on later in this section, and the nature of component functionality. In essence, components like object detection and prediction may exhibit poor performance when the ego vehicle is at a considerable distance from certain traffic objects, owing to limited sensor visibility. However, as the AD vehicle approaches these objects, it can still correctly identify or predict them, ultimately averting the violation. Therefore, we define the focus violation-inducing message as the last message that, if improved, could have prevented the AD vehicle from violating the rules. For instance, in the motivating example (Figure~\ref{fig:moti_example}), there exists a moment when, if the vehicle had consistently recognized the pedestrian correctly, the fatal crash could have been avoided. We contend that identifying such a critical frame can pinpoint the issue and significantly aid developers in diagnosing the root cause.

On the contrary, the planning component encounters challenges when it approaches traffic objects too closely. For instance, if the distance between the ego vehicle and other traffic objects is already less than the required emergency braking distance, it becomes impossible for the planning component to generate a safe trajectory, and a collision becomes inevitable. Therefore, we define the focus message for the planning component as the first message that violates the rules. We assert that planning should accurately respond to various situations as early as possible to ensure safety.

\textbf{Definition 11 (focus violation-inducing message).} The definition of focus planning violation-inducing message would be the first message that violates the driving specifications (detailed in \S\ref{design:message}). The focus violation-inducing message for the non-planning components is $\mathbb{M}[component_{VI}][index_{VI}]$ such that
\begin{equation}
\begin{aligned}
dtest(\mathbb{M}[component_{VI}][(index_{VI}): -1]) = \cmark,\\ and\; dtest(\mathbb{M}[component_{VI}][(index_{VI}+1): -1]) = \xmark
\end{aligned}
\end{equation}

\textbf{Overview.} Figure~\ref{fig:PVCA_overview} presents an overview of our workflow. Initially, we aim to rule out the possibility of planning to be the component responsible for the violation, considering its unique characteristics (explained in \S\ref{design:component}). If planning is determined to be the violation-inducing component, we perform violation checks for each planning output message, and the first message found to violate a specific driving policy is identified as the primary violation-inducing message. Conversely, we must investigate the potential of each other component causing the violation individually. Once the violation-inducing component is identified, we employ a binary-search approach to pinpoint the most recent message that, if operating correctly, could have prevented the violation.

\begin{algorithm}
\footnotesize
\caption{pseudocode code of the cause analysis algorithm }\label{algo:cause_localization}
\KwIn{A violation-inducing testing scenario S, the message collection $\mathbb{M}$ generated from the violation-inducing scenario with original AD software }
\KwOut{The violation-inducing component $component_{VI}$, and a focus violation-inducing output message $message_{VI}$}

\SetKwFunction{messageAttributionNonPlanning}{messageAttributionNonPlanning}
\SetKwProg{Fn}{Function}{:}{}
\label{algo:nonplanning}
\Fn{\messageAttributionNonPlanning{$component_{VI}$}}{
\tcp*[r]{Set the initial condition for the binary search}
    $left = 1$, $right = n$\;
    \While{$left + 1 < right$}{
    $mid = (left + right)/2$\;
        \If{$dtest(\mathbb{M}[component_{IV}][State_{mid}:-1]) = \cmark$}{
\tcp*[r]{Violation-inducing message is on the right side}
            $left = mid$\;
        }
        \Else{
\tcp*[r]{Violation-inducing message is on the left side}
          $right = mid$
        }
    }
    find a message $message_{VI}\in\mathbb{M}[component_{VI}][:]$ and belongs to state $State_{left}$\;
    \Return{$message_{VI}$}\;
}

\;
\SetKwFunction{messageAttributionPlanning}{messageAttributionPlanning}
\SetKwProg{Fn}{Function}{:}{}
\label{algo:planning}
\Fn{\messageAttributionPlanning{}}{
\tcp*[r]{Search for the 1st driving violation in planning output messages}
  \For {$m \in \mathbb{M}[planning][:]$}{
    \If{$violationTest(m) = True$}{
        \Return{m}\;
    }
  }
}

\tcp*[r]{Split the driving traces into a set of vehicle dynamic states }
$<State_1, State_2, ... State_{n}> = splitTrace(\mathbb{M})$\;
\tcp*[r]{We first check whether planning is the violation-inducing component}
\If{$dtest(\mathbb{M}[prediction+localization+control][:])=\xmark$}{
\label{algo:exclude}
  \Return{Planning, messageAttributionPlanning()}\;
}
\tcp*[r]{Exam whether component is violation-inducing one-by-one}
\For{$x \in \{ perception, prediction, control, localization\}$}{
    \If{ $dtest(\mathbb{M}[x][:]) = \cmark$}{
        \Return{x, messageAttributionNonPlanning(x)}\;
    }
}
\end{algorithm}

\subsection{Idealized Substitute Design for ADS Components}
\label{design:ideal_sub}
\begin{figure*}[t]
    \centering
    \includegraphics[width=\linewidth]{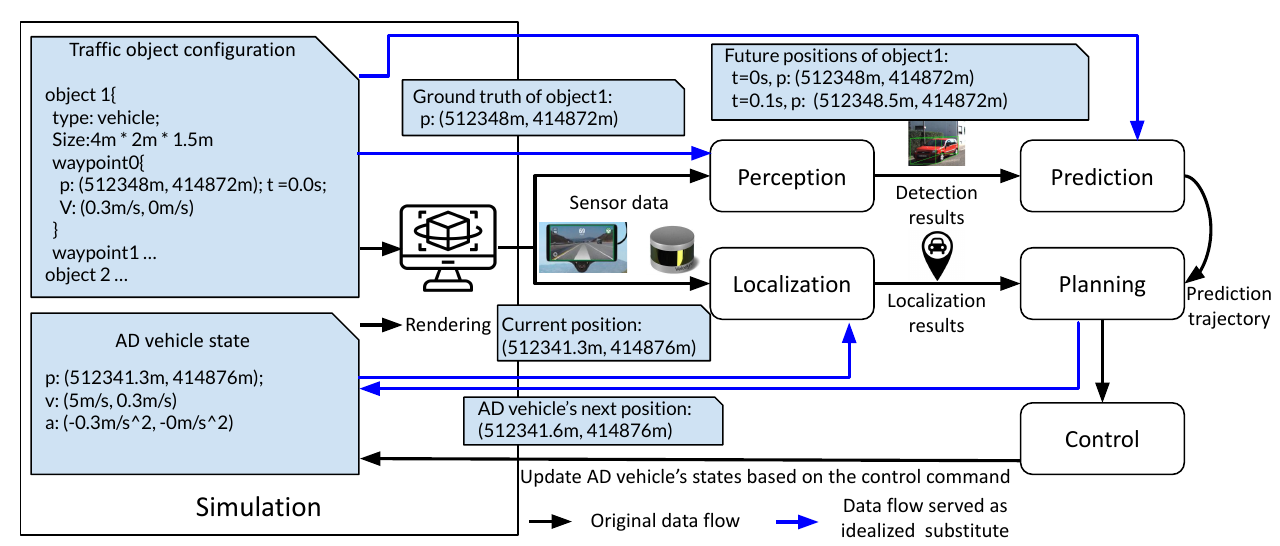}
    \caption{Illustration of the idealized substitute. \modify{Blue arrows are used to highlight idealized substitutes within the data flow. To provide a concrete example of this, we present a specific instance of the data flow, illustrating the information transmitted in each idealized substitute.}}
    \label{fig:sub_illustration}
\end{figure*}
To address \textbf{C2}, we designed idealized component substitutes for each component in the ADS. Our idealized component substitute should behave according to design expectations. In this work, we propose to extend the simulation bridge interface to provide counterfactual causality analysis for AD scenario-based testing.  

Typically, simulations interact with AD software using raw sensor data and control commands that are subsequently executed by actuators. However, we have identified that many intermediate results can be directly extracted or executed by the simulator itself. This capability allows us to design idealized substitutes for the perception, prediction, control, and localization components.

We present the design specifics of idealized substitutes for four components in Figure~\ref{fig:sub_illustration}. Original data flows are represented as black arrows. Originally, simulation should maintain two types of data internally. (1) \textit{traffic object configuration} is predefined before the simulation and encompasses information regarding the position of each object at each time. (2) \textit{AD vehicle state} is dynamically updated based on the simulation's control command results and includes real-time information about the ego vehicle's position, velocity, and acceleration within the simulated environment.

The nature of perception, prediction, and localization components is to understand the surrounding environment, which is all controlled by the simulator. We can directly set up the data flow such that the ground truth of each component output (already maintained in the \textit{traffic object configuration} and \textit{AD vehicle state}) is directly given (highlighted with blue arrows in Figure~\ref{fig:sub_illustration}).

In the case of the control component, which generates control commands based on a planning trajectory, the simulator employs motion models to estimate how the vehicle should move in response to the control commands. It then updates the vehicle's state within the simulation accordingly. For this purpose, we have adopted the sim-control technique from Apollo~\cite{apollo}, which directly updates the ego vehicle's states based on the planning trajectory. Since planning trajectories (the output message of the planning component) encompass information about how the vehicle is expected to operate in the future, this information can be employed to directly update the ego vehicle's position within the simulation. This approach is equivalent to assuming that the control component can generate control commands that perfectly align with the planning trajectory.
\begin{figure}
  \begin{center}
    \includegraphics[width=0.4\textwidth]{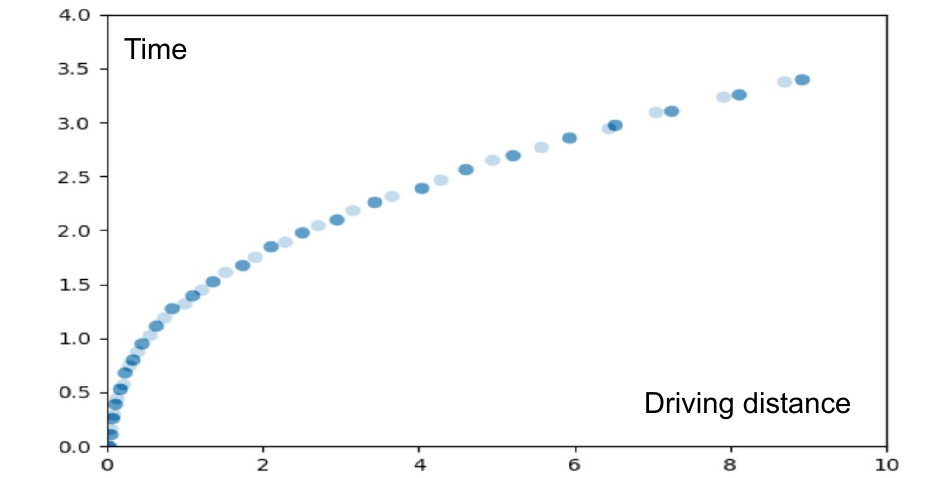}
  \end{center}
  \caption{Execution difference between two traces collected from the same testing scenario. Each node represents the driving distance and relative time when the planning component is executed.} \label{fig:exec_diff}
\end{figure}
\label{design:state}
\subsection{Component-Level Cause Attribution}
\label{design:component}
Even though we are able to substitute parts of the output messages with the idealized ones, we are still suffering from domain-specific limitations. A major limitation is the lack of idealized substitutes for the planning component. There is no idealized substitution for the planning component. As of now, we are not aware of any existing planning algorithm that can guarantee finding the most optimal planning solution within a limited reaction time~\cite{paden2016survey}. Furthermore, if we were to possess such a flawless planning component, there would be no need to test or debug it for potential errors.

Algorithm~\ref{algo:cause_localization} presents the detailed process of component-level cause attribution. To address this challenge, we must approach the planning component and non-planning components separately. The initial essential step involves ruling out the possibility that a fault in the planning component is the root cause. Adhering to the counterfactual causality principle, we can begin by examining the outcomes of $dtest(\mathbb{M}[localization\cup][:]\cup\mathbb{M}[prediction][:]\cup\mathbb{M}[localization][:])$ (line~\ref{algo:exclude} in Algorithm~\ref{algo:cause_localization}). If the violation persists even when idealized substitutes are employed for all other components, then planning is the root cause of the violation. Conversely, if the violation is averted, it suggests that planning can handle the testing scenario correctly, allowing us to exclude it as the root cause. In such a case, we will examine other components separately to attribute the cause.

\subsection{Vehicle Dynamic State-based Analysis}

Another challenge is that the execution trace under the same testing scenario cannot be precisely the same, so we cannot exactly replace the same message to perform causality analysis. As shown in Figure~\ref{fig:exec_diff}, the vehicle's position and time of each planning component execution cannot be matched, meaning that we cannot find the same message when rerunning the test.

To aid in identifying similar messages, we propose to cluster them based on the vehicle's dynamic characteristics, which encompass its position, velocity, and acceleration and serve to comprehensively represent a vehicle's dynamic state and encapsulate its current movement parameters. By grouping messages with similar vehicle dynamics, we can infer that they share similar inputs, such as sensor data and the ego vehicle's localization. Consequently, any faults or issues that arise are more likely to occur when these inputs are consistent or similar. This approach helps us identify and address problems more effectively by targeting specific clusters with shared vehicle dynamics.

For each recorded message, we can collect the vehicle dynamic and then calculate the vehicle dynamic state (we use "state" in the remainder of the paper):
\begin{equation}
    State(p, v, a) = p/p_{unit} || v/v_{unit} || a/a_{unit}
\end{equation}
where $p$, $v$, and $a$ are the position, velocity, and acceleration, respectively, of the ego vehicle and $p_{unit}$, $v_{unit}$, and $a_{unit}$ are pre-defined hyper-parameters. With this definition, we can cluster the messages into different states. 

For example, suppose that we have collected two driving records $T_A$ and $T_B$ under exactly the same driving-violation scenario configuration. In trace $T_A$, a planning message $M1$ is generated when ego vehicle is at position $(3m, 1m)$ and in trace $T_B$, a planning message $M2$ is generated at position $(3.1m, 1.05m)$. The ego vehicle's velocity and acceleration when generating $M1$ and $M2$ are exactly the same. The inputs of the two messages are very similar and if one message is violation-inducing in one trace, the other message is likely to be violation-inducing in the other trace. If we set $p_{uit} =0.2m$, we will have the same states for both messages. After applying state-based cause attribution, even though we can not directly replace the exact messages in the original driving-violation record, we are still able to replace messages which likely to induce the same fault and thus further attribute the cause on the message level. 

\subsection{Message-Level Cause Attribution}
\label{design:message}

We outline our message-level cause attribution algorithm in Algorithm~\ref{algo:cause_localization}. As detailed in the preceding subsection, our algorithm distinguishes between the planning component (line~\ref{algo:planning}) and non-planning (line~\ref{algo:nonplanning}) components since idealized substitutes are not available for the planning component. For the non-planning components, we employ binary search to gradually pinpoint the most recent message that could have prevented the AD vehicle from the driving violation if functioning correctly.

For the planning component, if we can exclude the possibility that other components are the cause of the driving violation, we could directly check the violation driving on planning outputs. Planning output by nature contains all necessary information including how the ego vehicle will operate at future moments to judge whether this output message could induce a violation. We also argue that this design is general and flexible. The developer can apply any driving based on the different operation domains and testing standards.
\section{Evaluation}
We aim to answer three research questions in the evaluation section:

\textbf{RQ1. Can the proposed cause localization technique effectively localize the cause? } 

\textbf{RQ2. How could the cause localization tool help analyze safety-violation results? }

\textbf{RQ3. What is the efficiency of our cause attribution tool?} 

\subsection{Experiment Setup}
We implement the cause attribution tool with Baidu Apollo 7.0, an industry-level open-source multi-component autonomous driving platform, and LGSVL, an industry-grade simulation for testing autonomous driving software. Due to the representativeness of system design, Baidu Apollo has become the default setup (all the related research chose Apollo as the platform to demonstrate their research except one paper used industry ADS provided by their industry partner~\cite{luo2021targeting}.) for related software engineering research~\cite{tian2022mosat, deng2022scenario, tang2021systematic, sun2022lawbreaker, tian2022generating, zhang2023building, cheng2023behavexplor}. For the idealized substitute of the perception component, we directly apply LGSVL's provided ground-truth mode. For the idealized substitute of the prediction component, we generate a ground-truth prediction trajectory from a safety-violation scenario configuration and publish it via a bridge between simulation and ADS. For the idealized substitute of the control component, we add another cyber node to extract the next position of the ego vehicle from the planning trajectory at run-time and implement a sensor plugin in LGSVL to directly update the vehicle state in simulation. The experiment is conducted on a machine with AMD Ryzen 9 3950X 16-Core and 128 GB of RAM.

\begin{table}
    \centering
    \begin{minipage}{\linewidth}
        \centering
        \footnotesize
        \caption{Benchmark created with real bugs. Pull requests (PR) and commits are from Github repo.}
        \label{tab:real_benchmark}
        \begin{tabular*}{\linewidth}{@{\extracolsep{\fill}} l|lll}
            \hline
            ID & Source & \begin{tabular}[c]{@{}l@{}}Module \\with fault\end{tabular} & Symptom \\
            \hline
            1  & PR 6826  & Planning & \begin{tabular}[c]{@{}l@{}}Dangerous\\nudging\end{tabular} \\
            \hline
            2  & PR 3634  & Perception & Collision \\
                        \hline

            3  & PR 3257  & Planning & \begin{tabular}[c]{@{}l@{}}Traffic line\\ violation\end{tabular} \\
                        \hline

            4  & PR 10706 & Planning & Speed violation \\
                        \hline

            5  & PlanFuzz V1 & Planning & \begin{tabular}[c]{@{}l@{}}Fail to normally\\ follow the lane \end{tabular}\\
                        \hline

            6  & PlanFuzz V7 & Planning & \begin{tabular}[c]{@{}l@{}}Fail to normally \\pass the stop \\sign intersection\end{tabular} \\
                        \hline

            7  & \begin{tabular}[c]{@{}l@{}}Commit\\ c803010\end{tabular}  & Planning &  \begin{tabular}[c]{@{}l@{}}Fail to stop\\ at destination\end{tabular}\\
                        \hline

            8  & \begin{tabular}[c]{@{}l@{}}Commit\\ cc890a4\end{tabular} & Planning & Collision \\
            \hline
        \end{tabular*}
    \end{minipage}
    \hfill
    \begin{minipage}{\linewidth}
        \centering
        \footnotesize
        \caption{Benchmark created with injected errors. 'CS' stands for collision scenarios. For each CS listed in the table, we created an instance under that collision scenario with the corresponding type of injected error.}         \label{tab:injected_benchmark}
        \begin{tabular*}{\linewidth}{l|l|l}
            \hline
            \textbf{Component} & \textbf{Injected error} & \textbf{CS} \\
            \hline
            \multirow{5}{*}{Perception} & Miss object detection result & 1, 2, 3, 4 \\
                                         & \begin{tabular}[c]{@{}l@{}}Wrong bounding\\ box estimation \end{tabular}& 4 \\
                                         & Wrong longitudinal distance & 1, 2, 3, 4 \\
                                         & Wrong lateral distance & 1, 2, 3, 4 \\
                                         & Wrong velocity measurement & 1, 2, 3, 4 \\
            \hline
            \multirow{2}{*}{Prediction}  & No prediction trajectory & 1 \\
                                         & Wrong prediction trajectory & 1, 2, 3, 4 \\
            \hline
            \multirow{3}{*}{Planning}    & Incorrect path planning & 3, 4, 5 \\
                                         & Incorrect speed planning & 1, 2, 3 \\
                                         & No planning trajectory & 1 - 5 \\
            \hline
            \multirow{2}{*}{Control}     & Wrong longitudinal command & 1, 2, 3, 4 \\
                                         & Wrong lateral command & 5 \\
            \hline
            Localization                 & Wrong lateral localization & 5 \\
            \hline
        \end{tabular*}
        \label{tab:benchmark}
    \end{minipage}
    
\end{table}

\textbf{Real bug benchmark.}  
To the best of our knowledge, there does not exist an ADS bug benchmark that includes ground truth fixes for those bugs. Therefore, to construct a benchmark for the evaluation of this work, we utilized bug-fix pull requests from GitHub as a ground truth for bugs and fixes. From a total of 9,711 merged pull requests, we first utilized keywords related to a driving violation (e.g., collision, violation) to filter related pull requests, then manually analyzed each bug fix to identify reproducible scenarios that might trigger driving violations. As a result, we were able to reconstruct 6 cases (ID 1, 2, 3, 4, 7, 8 in Table~\ref{tab:real_benchmark}). Also, we added 2 findings (ID 5, 6 in Table~\ref{tab:real_benchmark}) in prior work PlanFuzz~\cite{planfuzz} into our real bug datasets. Those vulnerabilities discovered in PlanFuzz~\cite{planfuzz} are also bugs that will lead to driving task violations. We do not include all of the discovered vulnerabilities since some of the consequences of the vulnerabilities are not severe driving violations in our scope (e.g., cannot change lanes).

\begin{figure}[t]
      \centering
          \includegraphics[width=1.\linewidth]{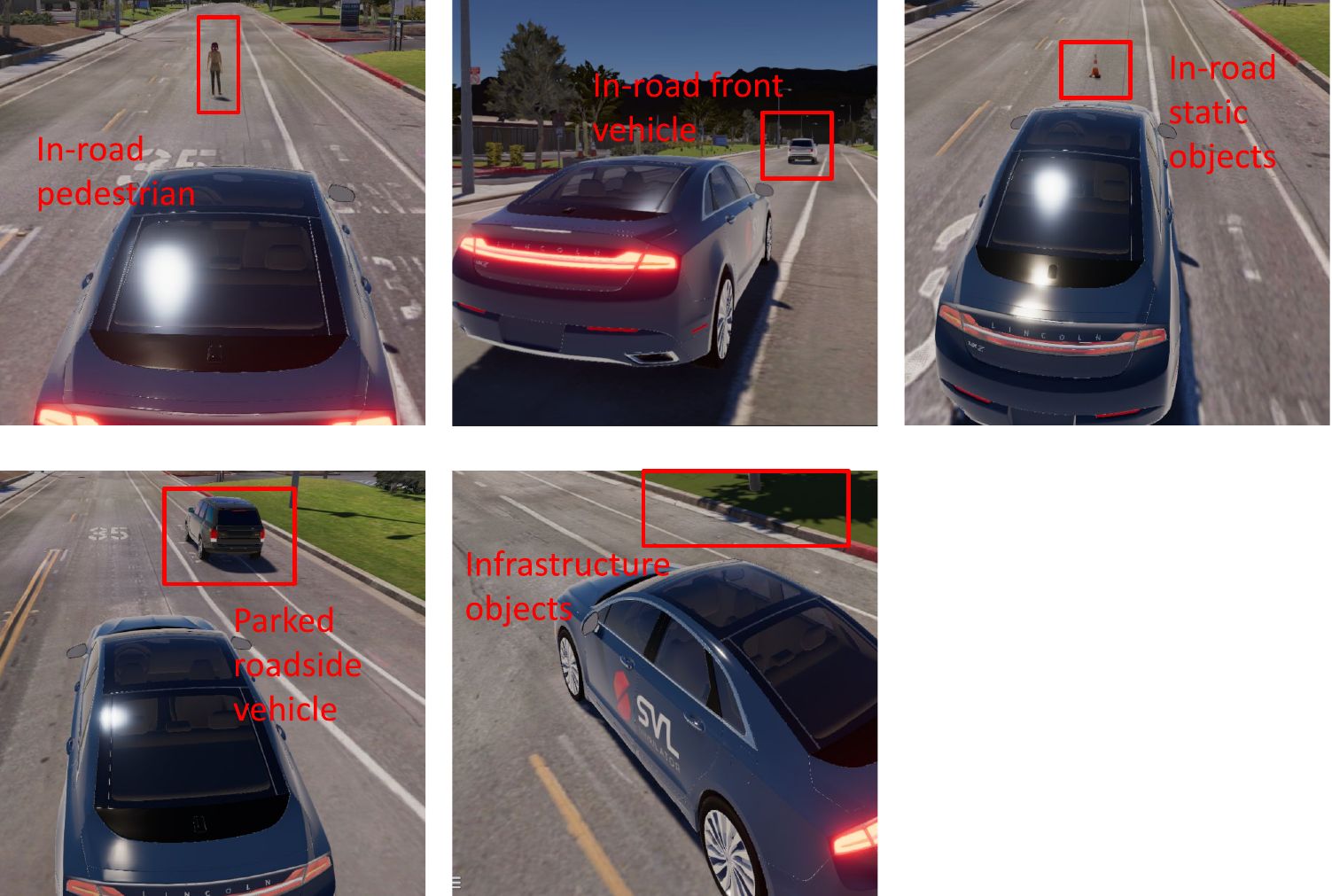}  
	\caption{\modify{Overview of 5 collision scenarios. From left to right: collision with in-road pedestrians; collision with in-road front vehicles; collision with in-road static objects; collision with parked roadside vehicles; collision with infrastructure objects (e.g., curb). The objects of the collision are highlighted with a red bounding box.} \alfred{see if we can annotate these in the figure to make it easier to understand when looking at the figure}}
		\label{fig:CS_overview}
\end{figure}

\textbf{Injected fault benchmark.} In the end, since we can only collect a limited number of real safety-violation bugs with the scenario, we created a benchmark with manually injected faults. We create a benchmark that includes possible faults for each component. First, we summarize 5 representative Collision Scenarios (CS) from NHTSA's L4 self-driving crash reports~\cite{NHSTA_report}: \textbf{CS1. } collision with in-road pedestrians; \textbf{CS2.} collision with in-road front vehicles; \textbf{CS3.} collision with in-road unmovable objects (e.g., traffic cone); \textbf{CS4.} Collision with parked road-side vehicles; \textbf{CS5.} collision with infrastructure objects (e.g., road divider) due to lane deviation. Fig.~\ref{fig:CS_overview} provides a snapshot for each collision scenario. 
We inject fault inside each component's output message to cause collision under each CS. Detailed information on our benchmark is listed in Table~\ref{tab:benchmark}. Note that in some cases, the injected error cannot lead to safety violations in all collision scenarios. For example, no matter how we change the object detection results, the ego vehicle will not drive out of the lane and cause a collision with roadside infrastructure since ADS can correctly recognize its position and the lane's position via the localization component and stored map information.


\subsection{RQ1. Effectiveness of Cause Attribution}
\textbf{Evaluation metric. }
For each instance in our benchmark, we conduct the cause attribution task with our tool. Our tool will report (1) the component which contains the error, and (2) one pair of inputs/output for that component that the developer should be focused on. 
Since each instance in the benchmark is created based on a real bug or injected faults, we can certainly tell whether the reported execution includes the bug/faults. We use two boolean metrics to report the effectiveness of the cause localization task: (1) component-level success: which is defined as whether the reported component is the component that contains a real bug/fault, and (2) message-level success: whether the reported output message is the result of the real bug/injected fault.


\begin{table}[ht]
\caption{\modify{Evaluation results on the benchmark with real bugs. \modify{The bug labeled with its ID in Table~\ref{tab:real_benchmark}. The letter 'Y' means the component/message level success is achieved.}}}
\label{tab:real_results}

\centering
\begin{adjustbox}{width=0.99\linewidth}
\begin{tabular}{lcccccccc}
\toprule
Bug ID & \multicolumn{1}{c}{1} & \multicolumn{1}{c}{2} & \multicolumn{1}{c}{3} & \multicolumn{1}{c}{4} & \multicolumn{1}{c}{5} & \multicolumn{1}{c}{6} & \multicolumn{1}{c}{7} & \multicolumn{1}{c}{8} \\
\midrule
\begin{tabular}[c]{@{}l@{}}Component\\-level success\end{tabular} & Y & Y & Y & Y & Y & Y & Y & Y \\
\hline
\begin{tabular}[c]{@{}l@{}}Message-level\\success\end{tabular}   & Y & Y & Y & Y & Y & Y & Y & Y \\
\hline

\begin{tabular}[c]{@{}l@{}}Reduction 
\\rate (\%) \end{tabular}    & 98.7 & 99.6 & 99.1 & 98.5 & 98.9 & 99.3 & 98.8 & 98.9 \\
\hline

Time (min)              & 3.1 & 17.2 & 2.2 & 3.8 & 3.1 & 2.9 & 2.4 & 3.0 \\
\bottomrule
\end{tabular}
\end{adjustbox}
\end{table}

\begin{table}[ht]
\tabcolsep 0.08in
\caption{Evaluation results on the benchmark with injected faults. \modify{'Prec.' = Perception, 'Pred.' = Prediction, 'Plan.' = Planning, 'Cont.' = Control, 'Loca.' = Location.} }
\label{tab:injected_results}
\centering
\begin{tabular}{lccccc}
\toprule
            Module          & \multicolumn{1}{c}{Perc.} & \multicolumn{1}{c}{Pred.} & \multicolumn{1}{c}{Plan.} & \multicolumn{1}{c}{Cont.} & \multicolumn{1}{c}{Loca.} \\
\midrule
\begin{tabular}[c]{@{}l@{}}Component-level\\ success rate\end{tabular} & 100\%  & 100\%  & 100\%  & 100\%  & 100\%  \\
\hline
\begin{tabular}[c]{@{}l@{}}Message-level\\ success rate \end{tabular}  & 100\%  & 100\%  & 100\%  & 80\%   & 100\%  \\
\hline
\begin{tabular}[c]{@{}l@{}}Avg. reduction\\ rate (\%) \end{tabular} & 99.89   & 99.91   & 99.87   & 78.7   & 98.63   \\
\hline
\begin{tabular}[c]{@{}l@{}}Avg. time (min) \end{tabular}          & 17.2   & 15.1   & 2.7    & 20.3   & 19.6   \\
\bottomrule
\end{tabular}
\end{table}

\textbf{Results.} We report the performance of our tool on the benchmark in Table~\ref{tab:injected_results} and Table~\ref{tab:real_results}. The results indicate that our tool is capable of achieving $100\%$ success in identifying which component is the cause of violations of system-level rules. In terms of message-level identification results, we observed that our tool is able to achieve perfect performance, except for one injected fault in control. For the injected fault, which our tool is not able to precisely the cause, we find that the reported output message is very close to the messages containing fault on the time dimension. The timestamp difference is smaller than 1s, meaning that our tool might still be useful if the developers also look into nearest messages. The potential cause of this failed case might be the randomness across different executions. We noticed that some of the vehicle dynamic states that exist in the original violation-included trace do not exist anymore when we replay the scenario again. Since we could not find the exact messages with the same vehicle dynamic states, we could only report the message with the most similar vehicle states, and that will lead to imprecise results. 

\begin{mdframed}[style=AnswerRQ]
\textbf{Answer to RQ1:} The evaluation results show that our tool can perfectly identify which component introduces the fault that leads to rule violations. Our tool is able to precisely one violation-introducing output message such that the developer only needs to focus on one execution of the violation-inducing component for most of the instances of our benchmark (>$98\%$).

\end{mdframed}

\subsection{RQ2. Reduction Rate of Cause Attribution}
\textbf{Evaluation Metric.} We employ the reduction rate as a metric to gauge the extent to which our automated tool can enhance the debugging process. The reduction rate is defined as $1-\frac{1}{|M|}$, where $M$ is the set of all messages in the violation-included driving traces. 

\textbf{Results.} The reduction rates are reported in Table~\ref{tab:real_results} and Table~\ref{tab:injected_results}. The reduction rate is at least $98.5\%$ on all real bugs and the average rate is inside the range of $98.7\% - 99.89\%$ for injected faults except for control. The reason that control has a much higher reduction rate is that the tool fails to correctly identify the violation-introducing message and we report a $0\%$ reduction rate for that specific instance since it cannot correctly reduce the debugging scope.

\begin{mdframed}[style=AnswerRQ]
\textbf{Answer to RQ2:} We showcase the substantial efficacy of our tool in minimizing the debugging scope within an Autonomous Driving (AD) context. When our tool successfully identifies the message responsible for the violation, it achieves a reduction of at least $98.5\%$. This reduction signifies that developers can now concentrate their efforts on addressing a single output message, eliminating the need to grapple with the intricacies of interdependencies and causality among hundreds of messages.
\end{mdframed}

\subsection{RQ3: Efficiency}
\textbf{Evaluation metric.} In this subsection, we evaluate the efficiency of our cause attribution tool on the benchmark. We use the running time as the metric.

\textbf{Results.} Our results show that on both the real bug benchmark and injected fault benchmark, our tool is able to achieve the fault isolation goal within 21 mins. In comparison, it is common for a researcher with experience in using AD software within our team to spend half or even one day to comprehend the inner workings of a single driving trace even with a mature data visualization tool. Also, surprisingly, we find that the running time of the cause attribution is much smaller if the fault/bug is located inside the planning component. The reason is that our tool will quickly determine whether the planning component is the root cause or not and then skip computation-intensive simulation tasks. Based on prior study~\cite{garcia2020comprehensive}, most of the bugs occur in the planning component, indicating that our tool performs particularly well in most bug cases.

\begin{mdframed}[style=AnswerRQ]
\textbf{Answer to RQ3:} On the benchmark created with real bugs, our tool takes at most $17.2$mins to complete the cause attribution task and at most $3.1$mins to complete the task if the bug is inside the planning component. Compared with manually analyzing the data at each moment, our tool is able to greatly save developers time on bug fixing.
\end{mdframed}

\section{Case Study}
\modify{Besides the evaluation section, we also created a case study to demonstrate how the automated cause analysis could help the developers.} 

\modify{In one specific case, illustrated by PlanFuzz V1~\cite{planfuzz}, our automated tool demonstrated its remarkable capability to streamline the debugging process. It successfully narrowed down the initial debugging scope from a daunting four hundred thousand lines of code to a mere 53 lines of code. This significant reduction in scope was instrumental in identifying the root cause of the issue. The underlying problem stemmed from the inappropriate configuration of a crucial variable, namely, \textit{obstacle\_lat\_buffer}. This variable dictates how the autonomous driving (AD) vehicle should respond to static obstacles, and its static configuration led to the issue at hand. Notably, the variable lacked dynamic adjustments. As a result, the AD vehicle invariably came to a halt when encountering two entirely off-lane obstacles, with no adaptability in its response strategy.}

\textbf{Detailed approach.} With the ability provided by our cause analysis tool, we can reduce the scenario-level testing oracle into the execution-level oracles, and thus we can extract one failed planning test execution, referred to as failed execution. Meanwhile, we could also use the testing tool PlanFuzz~\cite{planfuzz} to extract a planning execution without any violation (before the testing tool successfully discovers the vulnerability, it will generate similar test cases without any violation). We referred to the planning execution without violation as the passed execution. We collected the code coverage of each line in the source code separately. We use the functions $passed(l)$ and $failed(l)$ to denote whether a specific line $l$ is covered in passed execution or failed execution. We also denote $total_pass$ and $total_fail$ to be the total number of lines executed in passed or failed execution. We followed the Tarantula~\cite{jones2005empirical} technique and ranked each line with the following suspiciousness scores:
\begin{equation}
    suspiciousness(l) = 1 - \frac{\frac{failed(l)}{total\_failed}}{\frac{failed(l)}{total\_failed} + \frac{passed(l)}{total\_passed}}
\end{equation}

\textbf{Results.} In the end, with the suspiciousness score, we can filter 53 lines of code with the highest suspiciousness score since they are only covered in the failed execution. Among 53 lines of code, the root cause of the violation (e.g., the code that checks whether the remaining space is enough for the AV to safely bypass) is included.

\section{Related work}



\noindent\textbf{Debugging in ADS.}
Even though lots of research efforts have been devoted to designing and developing effective and efficient testing frameworks to discover safety hazards in autonomous driving systems, only a few works focus on the automated tools for post-violation-discovery tasks after discovering safety hazards, such as cause analysis, and bug repairs. 

Stein et al.~\cite{von2021icra} proposed a technique to automatically reduce the environment for debugging robotic systems, which is the most related to our work. Similar to the idea of delta debugging~\cite{zeller2002simplifying}, the technique aims to perform a partition of the environment space and finds a minimal reduced subset of the environment that can still lead to the same failure of the system. Even though such environment reduction can reduce developers' efforts, it treats the whole system as a black box and cannot provide any insight into which part of the program caused the system failure.

Another related line of research is fault detection in robotic systems~\cite{khalastchi2018fault}. Model-based and data-driven approaches~\cite{khalastchi2018fault} have been designed to monitor the components of robotic systems at runtime. However, building a perfect oracle to determine the existence of faults for components in robotic systems is still an open problem due to the complexity and noise in such systems~\cite{he2019system}. To the best of our knowledge, there is no existing concrete design or implementation that can detect the faults in multi-component autonomous driving systems. Moreover, such a fault detection method only focuses on one component of the system and thus cannot infer the interaction between components and cannot use causality to infer the cause of system failure.

A recent work~\cite{yamaguchi2020specification} proposed to use the formal specification to identify which component is more likely to have faults. However, writing specifications is challenging and time-consuming. Moreover, it cannot provide a counterfactual relationship between the driving violation and the messages of each ADS component.

In general, the key difference between our work and prior works is that we are aiming to identify the causality relationship between driving violation and each ADS component and each ADS component output (message). This can bring two benefits: (1) We can directly tell which components/messages are violation-inducing. (2) We do not rely on constructing specifications or testing oracles, which is challenging in CPS area~\cite{yamaguchi2020specification, khalastchi2018fault}.

\noindent\textbf{Cause Analysis in Software Systems}
Root cause analysis has been studied widely in other areas. Liu et al.~\cite{liu2021microhecl} dynamically construct a service call graph to analyze propagation chains for availability issues in large-scale microservice systems. Malik et al.~\cite{malik2013automatic} apply machine learning techniques to analyze the root cause of performance deviations in load testing of large-scale systems. Murali et al.~\cite{murali2021scalable} proposed using data mining to learn from past reports to find the root causes of symptoms. None of the aforementioned works apply counterfactual causality to attribute the cause. 

A closely related work that also applies counterfactual to find the root cause of defects is Causal Testing~\cite{johnson2020causal}. However, it aims to infer the relationship between the inputs and behavioral changes in the program. In our work, however, we are aiming to infer the relationship between driving violations and messages, which are intermediate computation results of software components. This unique challenge requires our idealized component substitute design, which makes prior work non-applicable here. 

\section{Threats to validity}

\textbf{Representativeness of the benchmark with injected faults.} First, construction of the injected faults involves manual efforts (manually determining how to inject and the triggering condition of the faults), which may be subjective. To reduce the threat, we try to minimize the value of injected error and narrow down the triggering conditions. Besides, we tried to be complete in terms of which data field in messages could be wrong and the safety-violation scenarios. We also argue that this is the best effort we can make. There exists some work studying injecting faults into ADS~\cite{jha2018avfi, jha2019kayotee, jha2019ml}. However, their focus is on how to cause system-level failure with injected faults instead of making the faults realistic. We think how to inject faults in AD systems is a research topic worth studying in the future, which can benefit research works on cause analysis, fault localization, and bug repair in this area. 

\textbf{Multiple independent violation-inducing ADS components.} We acknowledge one limitation of our work, which is our assumption that there is only one violation-inducing ADS component. We believe that this assumption is relatively unlikely today. Currently, autonomous driving companies can achieve long distances (up to 18K miles for Waymo) without disengagements~\cite{waymo_18k}, implying that the probability of a particular bug being triggered within 1 mile is less than $\frac{1}{18,000}$. The probability of two independent root causes leading to the same driving violation would be even smaller, approaching $\frac{1}{18,000^2}$, which is significantly less likely than a single root cause. We consider this problem as an interesting working direction to explore in the future.



\textbf{Randomness in ADS testing.} We recognize that randomness in scenario-based testing could potentially affect the validity of our results. To address this concern, we have implemented vehicle dynamic state-based analysis to reduce discrepancies between driving traces, even when using the same driving scenario. We propose that this randomness can be further mitigated by running the scenario multiple times and aggregating the testing results from different runs, a concept that can be easily integrated into our DVCA framework. We view the measurement and mitigation of randomness as a promising avenue for future research.

\textbf{Limitations of the real bug benchmark.} So far, we are not aware of any existing bug dataset/cause localization benchmark for driving violations in the AD context. Lacking of good benchmark is a common research challenge in cyber-physical systems in general. Recent work in such domain is only able to evaluate their tool on 8 real-life bugs~\cite{he2019system}. We are only able to reconstruct a very limited number of driving violation scenarios for the following reasons: (1). Not all bugs reported in the pull request are bugs that can lead to driving violation behavior in AD systems. The bugs related to the build, documentation, and UI interface are out of the scope of our paper. (2). In almost all the pull requests on Github, there is no clear information about the explanation of the bug, how to trigger this bug, and what's the symptoms of this bug. We have to manually understand the cause of the bug, and how to create a scenario to trigger this bug and eventually lead to a driving violation. This process is naturally slow and leads to a very low success rate in reconstructing the safety-violation scenarios since we have very limited information compared with original AD developers.
\section{conclusion}
In this work, we propose a novel driving violation cause analysis (DVCA) tool. We design idealized component substitutes to enable counterfactual analysis of ADS components by leveraging the unique opportunity provided by the simulation. We evaluate our tool on a benchmark with real bugs and injected faults. The results show that our tool can achieve perfect component-level attribution accuracy (100\%) and almost (>98\%) perfect message-level accuracy. Our tool can reduce the debugging scope from hundreds of complicated interdependent messages to one single computation result generated by one component. Our tool underscores the debugging opportunities that simulation can provide for CPS software. We also delve into the challenges and prospects in the realm of ADS cause attribution. We aim to draw increased attention to the analysis of causes in ADS, given the rapid deployment speed of ADS.


\bibliographystyle{IEEEtran}{
\bibliography{./bib/general.bib}
}
\end{document}